\documentstyle[11pt,aaspp4]{article}
\slugcomment{\it (To appear in Astronomical Journal, September 1998)}

\begin{document}

\title{Luminosity Functions and Evolution of Blue
       Galaxies In A Deep Multicolor CCD Field Survey}

\author{Charles T. Liu\altaffilmark{1} and Richard F. Green}
\affil{National Optical Astronomy Observatories, Tucson, AZ 85726}
\author{Patrick B. Hall\altaffilmark{2}}
\affil{Steward Observatory, University of Arizona, Tucson, AZ  85721}
\and
\author{Patrick S. Osmer}
\affil{The Ohio State University, Columbus, OH 43210}

\altaffiltext{1}{Current address: Department of Astronomy, Columbia 
University, 550 E 120 St, New York, NY  10027.}
\altaffiltext{2}{Currently at the Department of Astronomy, University
of Toronto.}

\begin{abstract}

A complete sample of 659 field galaxies with 
17.0$<$U$\leq$21.1, each with U-B-V-R-I7500-I8600
photometry, has been selected from a deep field survey 
which covers 0.83 deg$^2$ along six lines of 
sight (Hall et al. 1996a).  Each galaxy's
spectral type and redshift has been estimated
using a multicolor photometric technique (Liu \& Green 1998).
Total number counts of the galaxies in the U-band give a count slope 
d(logN)/dM = 0.55 $\pm$ 0.05, consistent with previous studies.

The 545 galaxies in the sample classified
as spectral type Sbc or bluer are analyzed for signs
of evolution with redshift, and for unusual star formation histories.
The U-band luminosity function of these blue galaxies at 0.02$<$z$<$0.15 
has a steep $\alpha$ $\simeq$ -1.85 down to M(B)$\simeq$-14.
The luminosity functions at 0.15$\leq$z$<$0.3 and 0.3$\leq$z$\leq$0.5 
show significant evolution in M* and $\phi*$,
at levels consistent with those found in the Canada-France and
Autofib Redshift Surveys.

A significant population of very blue 
(rest frame U-B$ < -$0.35) galaxies, with spectral energy 
distributions indicating strong starburst activity, 
is observed at z$\gtrsim$0.3 but not at z$<$0.3.  This population is
confirmed via spectroscopy of part of the sample.
These may be galaxies 
temporarily brightened by global starbursts, which subsequently 
fade and redden at lower redshifts.

\end{abstract}

%\keywords{globular clusters,peanut clusters,bosons,bozos}

%%sI
\section{Introduction}

The evolution of field galaxies with redshift is a
phenomenon that has been intensely scrutinized, especially recently.
It is a subject of debate at every level, including the most basic:
what is the zero-redshift galaxy luminosity function?  
The common approach to this question 
is to model the luminosity function (LF)
using the parameterization of Schechter (1976), which contains 
a characteristic luminosity M*, a galaxy space density parameter $\phi$*,
and a faint luminosity end with power-law slope $\alpha$.  Thanks to
the work of numerous authors on samples of galaxies now totaling
upwards of 10$^4$ objects,
a reasonably consistent picture is gradually forming
regarding at least two facets of the local LF -- with the understanding,
of course, that many important details have yet to be reconciled. 
First, $\alpha$ is probably roughly
$-1.0 \pm 0.3$ for the sample of all galaxies in the local universe
(see, e.g., Loveday et al. 1992; Marzke et al. 1994a,1994b; da Costa et al.
1994; Lin et al. 1996).  Second, the
luminosity functions of blue galaxies, emission-line galaxies
and morphologically later-type galaxies tend to have steeper $\alpha$ than
redder, non-emitting and earlier-type
ones, with estimates ranging from $+$0.2 (Loveday et al. 1992) for early-type
galaxies, to $-$0.9 for galaxies with [OII]$\lambda$3727 $\geq$ 5\AA\ 
(Lin et al. 1996), to $-$1.87 for morphologically identified Sm-Im galaxies
(Marzke et al. 1994b).

As we look at more field galaxy samples with fainter apparent magnitude 
limits, the key question becomes: how does the luminosity function, and
by inference the field galaxy population, evolve with redshift?
In the absence of redshift information, faint galaxy counts have been
the primary tool for measuring evolution.  
The case for strong evolution in the field galaxy luminosity function was
made, among others, by Maddox et al. (1990), who used number count data 
-- specifically,
a steep number count slope dlogN/dM -- to show that significant evolution may 
be occuring in galaxies at 17$<$B$<$21.
Multicolor surveys such as those of Koo (1986) and Jones et al. (1991)
have added dimensions to number count studies by examining the color
distributions of galaxies over long wavelength and magnitude baselines.
The consensus appears to be that dlogN/dM steepens with decreasing
wavelength, suggesting stronger evolution in blue galaxies than red ones.
Koo (1986) also used his multicolor data to estimate redshifts in his
galaxy population -- the so-called photometric redshift technique, which
has since been extensively developed -- and concluded that an excess of star
formation is present in the field galaxy population by $z \sim$ 0.4.

Ideally, redshifts for all the galaxies in a survey should be obtained
when studying the evolution of those galaxies, 
because they provide critical information about the
distances and absolute magnitudes of each object.
Broadhurst, Ellis \& Shanks (1988) and Colless et al (1990; 1993) suggested
that evolution of the field galaxy luminosity function, probably
in the form of increased number density with redshift, was necessary
to explain the redshift distribution of their magnitude-limited,
20 $\leq$ B$_J \leq$ 22.5 survey.  Other redshift surveys by Lilly, Cowie
\& Gardner (1991), Songaila et al. (1994) and Cowie et al. (1996)
have concluded that little or no luminosity evolution occurs from
$z$ = 0 to $z \sim$ 1, but suggest the appearance of an additional,
starbursting population of objects at high redshift which 
have since faded -- in essence, another sort of number density evolution.

Several recent studies have broken new ground in the study of faint
galaxy luminosity functions, by obtaining redshifts for $\sim$10$^{3}$
objects in complete samples of faint galaxies.  The Canada-France 
Redshift Survey (Lilly et al. 1995, hereafter CFRS VI) used
an I-band selection criterion to study 730 galaxies with median redshift
$<z>$ = 0.56, and found strong, uniform brightening of the luminosity function
among the blue (i.e., with colors of a template Sbc galaxy or bluer) galaxy
population.
The Autofib Redshift Survey (Ellis et al. 1996; Heyl et al. 1997) 
assembled over 1700 redshifts
from a number of B-band selected samples to study an 
impressively large magnitude
range of 11.5 $\lesssim$ B$_J \lesssim$ 24.  
They also found significant evolution of
the blue galaxy luminosity function; the faint-end slope of their Schechter
function parameterization was a steep $\alpha \sim -$1.5 in their medium
and high-redshift bins (0.15 $\leq z \leq$ 0.35 and
0.35 $\leq z \leq$ 0.75, respectively).
Most recently, the LF of field galaxies in the CNOC1 Redshift Survey
has been analyzed by Lin et al. (1997), who also found that objects with
typical Sbc galaxy colors or bluer have a steep $\alpha \simeq -$1.4.

Deep spectroscopic surveys such as these are extremely powerful datasets
with which galaxy evolution can be studied in detail.  
An ideal faint galaxy survey, then, would combine faint limits, large numbers,
multicolor photometry and redshifts to look at the evolution of a complete
galaxy sample from many directions -- luminosity functions, number counts,
and spectrophotometric distributions.  The primary limiting factor for
obtaining such a sample is the enormous amount of telescope time required
to measure redshifts with spectroscopy; if reliable redshifts for the survey
galaxies could be obtained with colors alone, such a project becomes much
more observationally feasible.

In this work, we present the results from a complete U-band selected sample of
galaxies from a deep, multicolor CCD survey in six optical and near-infrared
passbands.  We use a photometric
technique to estimate the redshift and spectral type of each
galaxy.  With those data, we analyze in deatil the luminosity
function and spectrophotometric properties of the 545 galaxies in the
sample with spectral type Sbc or bluer.  The multicolor
photometry we have measured for each galaxy gives us additional leverage:
we minimize possible systematic errors from K-corrections, and
we can examine the relative contributions of different-colored galaxies
at different redshifts to the observed galaxy population.  Our primary 
goal is to examine the evolution of field galaxies; particularly, we
will address the star formation histories of blue field galaxies, to
which our U-band selected sample is more sensitive than samples chosen
with longer-wavelength passbands.  Throughout
this paper we use H$_{\rm o}$ = 100$h$ km/s/Mpc ($h$ = 0.8), 
and q$_{\rm o}$=0.

\section{Sample Selection and Galaxy Number Counts}

Our sample has been assembled from the Deep Multicolor Survey (DMS)
of Hall et al. (1996, hereafter HOGPW).  The extracted dataset is
a complete, magnitude-limited sample of
659 galaxies of all types, with 17.0$\leq$U$\leq$21.1.
All these galaxies have photometric accuracies of
$\sim$10\% photometry or better in each of the six
DMS passbands U, B, V, R, I75 and I86.  The procedures 
used for galaxy selection and
photometry of the survey galaxies are detailed in Appendix I.

The basic yet informative technique of galaxy number counts is easily
applied to our final, complete U-selected dataset, and we do so here.
The U-band number counts for all DMS galaxies is shown in Figure 1;
the magnitude limit for our sample, U$\leq$21.1,
is marked by the dotted line.
We present the plot of the number of galaxies per square degree vs. 
apparent U magnitude in Figure 2.  Also plotted for comparison are 
number counts from the photographic U-band surveys of 
Jones et al. (1991), which cover the magnitude range 18$<$U$<$21,
and Koo (1986), which cover U$>$19.
Those two surveys agree at U$\simeq$21, but are about a factor of
two apart at U$\simeq$19.  (Jones et al. attributed the
difference to local large-scale structure.)  Our absolute number
counts agree very well with Koo (1986), suggesting that the mean
field density of the lines of sight covered by the DMS is similar
to the SA57 and SA68 fields observed by Koo; that is, we are not
likely to be strongly contaminated by clustering at z$\leq$0.5.

We can compute the slope of the number counts with magnitude, dlogN/dM, 
using an error-weighted least squares fit algorithm; we derive
dlogN/dM = 0.55 $\pm$ 0.05, an intermediate value between 
the Jones et al. (1991) and Koo (1986) measurements of 0.49 
and 0.68 respectively.  These consistency checks give us confidence
that our sample is indeed complete and unbiased by significant 
selection effects or systematic errors.

\section{Photometric Measurement of Redshifts and Spectral Types}

The application of the photometric techniques developed in 
Liu \& Green (1998, hereafter LG98) to all the galaxies, in order to
measure their redshifts and spectral types, is a key step in our
analysis of the dataset.  We refer the reader to LG98 for a 
thorough discussion of the method, and present only a brief summary here.

The photometric redshift/typing system we use is grounded on the basic
principles first shown by Baum (1962) and later developed further by
Koo (1985), Loh \& Spillar (1986), Connolly et al. (1995) and others.
Essentially, all galaxies not dominated by active galactic nuclei have
spectral energy distributions (SEDs) 
in the near-UV to near-IR wavelength range
distinctive enough that, even as they are redshifted into redder
observed passbands, their SEDs can be distinguished from galaxies of
other types and/or other redshifts with a high level of accuracy with
broad-band colors alone, given sufficient wavelength coverage.
Although a
photometrically determined redshift cannot compare to spectroscopy
for measurements of individual galaxies, it can be an excellent statistical
tool to examine galaxy populations in large surveys, for which 
obtaining spectroscopic redshifts is by far the most difficult and
expensive in terms of telescope time.

The photometric method of LG98 uses the UBVRI75I86 data to place
galaxies into five spectroscopic categories: E/S0 galaxies (little
or no star formation), Sab (weak star formation), 
Sbc (active star formation), Scd (strongly
active star formation), and irregular (starburst).  Simultaneously,
the method allows
us to estimate galaxy redshifts to 1$\sigma$ accuracies 
of $\Delta z \sim$ 0.05, and spectral types to within $\pm$1 galaxy
spectral type, in the redshift range 0 $< z <$ 1.  This is essentially the
same precision achieved by previous authors with UJFN photographic
photometry (e.g., Koo 1985; Connolly et al. 1995) and 
multicolor photoelectric and CCD photometry (e.g., Loh \& Spillar 1986).
Since the six-filter dataset slightly ``overdetermines'' the 
principal components of most galaxies (Connolly et al. 1995), we can
avoid using the apparent magnitude of a galaxy in our redshift 
determinations.  Furthermore, the division of the standard
I-band into two narrower bandpasses (I75 and I86) gives us more leverage
than the broad I or photographic I$_N$ at
higher redshifts, as prominent spectral features such as H$\alpha$ 
(at z $\sim$ 0.3) and the 4000 \AA\ break (at z $\sim$ 0.9)
are redshifted into that wavelength range.  Finally, LG98 uses an
algorithm based on empirical data, so incompleteness problems caused by
``negative redshift'' determinations, which occasionally happen
for the Connolly et al. (1995) technique, do not occur.

As discussed in LG98, we emphasize again the difference between the 
$spectral$ type of a galaxy and its $morphological$
type.  Although there is generally good correspondence between
the SED of a galaxy and its position on the Hubble sequence, there
are many exceptions to this rule, especially among peculiar galaxies,
strongly interacting systems or mergers (see, e.g., Kennicutt 1992;
Liu \& Kennicutt 1995).  In our analysis, we use the derived
galaxy spectral types only as 
a sequence of relative star formation rates and broad-band colors.
Except where noted otherwise, we use in this work the terms ``spectral type''
and ``galaxy type'' to describe not the morphology of a given galaxy, 
but rather spectrophotometric properties.

The application of the redshift-type identification algorithm is
summarized in Appendix I.  Of the 659 galaxies, 104 were classified
as E/S0 or Sab galaxies.
It has been reasonably well established 
(see, e.g., Lilly et al. 1995; Ellis et al. 1996;
Heyl et al. 1997; Lin et al. 1997) that field ellipticals and early-type
spirals evolve little if at all at redshifts less than unity,
whereas galaxies with Sbc colors or
bluer are the primary contributors to the evolution of the field
galaxy population. Since the primary goal of this work is to
examine field galaxy evolution, we therefore set aside this
subset of early-type galaxies and
concentrate our efforts on the rest of the galaxy population --- namely,
the 545 galaxies in the sample that have been classified as
Sbc, Scd or Irr (starburst) galaxies.  All subsequent analysis 
we present in this paper is conducted on the survey galaxies 
classified as Sbc or bluer, and we refer to them 
collectively as the blue galaxy sample.

\section{Redshift Dependent Luminosity Functions}

Constructing the luminosity function of these blue galaxies as 
a function of redshift is perhaps the best way of quantifying
evolution in the blue galaxy population.
With our data, an added complication must be taken into account
when computing the absolute magnitude of each galaxy: the
redshift determined has a large enough error ($\sigma$ = 0.05)
to affect the luminosity calculation significantly.  The method
we use to compensate is to treat each galaxy as if it had a
probability-weighted distribution in redshift.  A similar
conceptualization of the problem is dicussed in 
SubbaRao et al. (1996); using the
UJFN-derived photometric redshifts of Connolly et al. (1995)
and a modified version of the C-method (Lynden-Bell 1971),
they are able to reproduce a spectroscopic-redshift luminosity
function rather well.  In this work we adopt a modified version
of the 1/V$_{max}$ formalism (e.g., Schmidt \& Green 1986) to
compute our redshift-dependent luminosity functions.

\subsection{The Probability-Smoothed Luminosity Distribution}

Consider a galaxy with an apparent magnitude $m_f$ in a passband
$f$, and redshift
$z \pm \sigma$.  If $\sigma$=0, then the absolute magnitude is
simply 

\centerline{ $M_f$ = $m_f -$ 5 log(d$_L(z)$) $-$ 25.0 $- k(z)$ }

\noindent
where d$_L(z)$ is the luminosity distance in megaparsecs, and
$k(z)$ is the K-correction in that passband for the spectral 
energy distribution of the galaxy.  The contribution of that
galaxy to the luminosity distribution is then a delta 
function of amplitude unity at redshift $z$.  

In the case where $\sigma >$ 0,
and the error distribution is Gaussian, the galaxy
can be thought of as adding a series of
fractional contributions to the luminosity distribution
in the redshift space surrounding $z$.  Such a fraction at,
for example, redshift $z + \Delta z$ and with a
differential redshift width $dz$, would have an absolute magnitude

\centerline{ $M_f^{\arcmin}$ = $m_f -$ 5 log(d$_L(z + \Delta z)$) $-$ 25.0 $- k(z + \Delta z)$}

\noindent
and have an amplitude

\centerline{ N$_{z + \Delta z}$ = 
P$_G(z + \Delta z,z,\sigma) dz$ / A$_G(z + \Delta z,z,\sigma)$}

\noindent
where P$_G$ and A$_G$ are the Gaussian probability function and its integral,
respectively (see, e.g., Bevington \& Robinson 1992).  

This ``fuzzing'' of a galaxy's luminosity distribution 
in redshift space is straightforwardly achieved 
numerically, with a choice of $dz << \sigma$ to minimize random magnitude
errors.  For our dataset, $\sigma$ = 0.05; our choice of $dz$ = 10$^{-3}$.
Our computational algorithm divides each galaxy into a Gaussian-weighted
luminosity distribution with 300 bins, from $z - $3$\sigma$ to
$z + $3$\sigma$ (i.e. $z \pm$ 0.15).  The entire distribution for each 
galaxy is normalized to unity.  

With the methodology described above, in some cases the luminosity distance 
is computed with redshifts that are close 
to zero.  As noted by SubbaRao et al. (1996), this can contribute large
systematic errors, because near $z$ = 0 a small variation in redshift can
imply an enormous change in distance modulus and in the accesible 
volume V$_{max}$.  We therefore consider any
fractional luminosity contributions with $z <$ 0.02 as sampling the volume
within that radius, and make $z$ = 0.02 our low redshift cutoff when
computing the luminosity function in the low-redshift bin.  In any case,
galaxies within that volume, corresponding to a recessional velocity
$cz <$ 6000 km/s, are subject to systematic effects from local large-scale
structure and the local supercluster; accounting properly for those effects
would be a task in itself, and does not lie within the scope of this work.

Our multicolor survey gives us an additional measure of accuracy when
computing the absolute magnitudes of our galaxy.  As redshift increases,
the accuracy of the magnitude measurement becomes increasingly 
affected by the inaccuracy of the K-correction -- in other words, on
the precision with which the SED of the galaxy is actually followed 
by the SED fitted to it. We can reduce our dependency on this parameter
by computing the absolute magnitude of one rest-frame
passband using the flux through
a observed redder passband for high redshift objects (cf. CFRS VI).
As an example, the rest effective wavelength
($\lambda_{eff}$) of our U-filter is the same as the observed
$\lambda_{eff}$ of our B-filter at $z$ = 0.200.
So in the range 0.15$>z>$0.40, we use the flux through the B-filter,
normalized to the rest-frame U-band flux expected for this galaxy, and
the appropriate K-correction to the B-filter to compute the absolute
U magnitude.  
In this way, we measure the rest-frame 
U-band wavelength range much more directly than if we had to rely on a large
K-correction.

\subsection{The Modified 1/V$_{max}$ Method}

In the standard 1/V$_{max}$ method, each galaxy contributes a weight
to the luminosity function equal to the inverse of the accessible
volume within which it can be observed.  The accessible volume,
referred to here as V$_{max}$, is the total comoving volume within
the redshift boundaries of the sample, where the given galaxy could
be and fall within the selection criteria of the sample.  In our case,
the relevant criterion is that its apparent U magnitude 
lies between 17.0 and 21.1.  The available volume is based on the
effective solid angle of 0.830 square degree covered by the survey.
A disadvantage of the 1/V$_{max}$ method for deriving a spatially 
smoothed luminosity function is that it is sensitive
to clustering within the volume; our sample, however,
is not likely to be significantly contaminated by
clustering, as our number counts above have shown.

In the case of a probability-weighted luminosity distribution for
individual objects, it
is straightforward to compute V$_{max}$ for each fractional galaxy;
correspondingly, its contribution to the luminosity
function is (1/V$_{max}$) $\times$ N$_{z + \Delta z}$.  Assembling
the luminosity function (LF) is then a matter of summing those contributions
within absolute magnitude bins.  The error in each magnitude bin was
estimated with a standard bootstrap technique, using 200 random samplings
(with duplication allowed) of the observed dataset.

\subsection{Simulations}

The obvious systematic error that comes from creating a luminosity
function with ``fuzzy'' redshifts is that objects near the peak
of the distribution will have some part of their light distributed
toward brighter magnitudes.  Similarly, galaxies toward the bright
and faint ends of the absolute magnitude distribution will have
some part of their partial contributions scattered into regions
which otherwise would have few or no galaxies.  This effect will
cause us to underestimate those parts of the LF
that contribute the most light, and overestimate those parts which
contribute the least.  Another systematic error which can arise 
is that at low redshifts ($z \sim 0.05$), the lower-redshift side of
the redshift probability curve contributes significantly more to 
the luminosity function than the higher-redshift side; this
produces a bias toward lower redshifts and luminoisities.

To correct for these systematics, we have conducted extensive
numerical simulations to quantify the effects of using fuzzy redshifts
to derive galaxy luminosity functions.  In the ranges $-21 < M* < -18$,
$0.005 < \phi* < 0.05$ and $-0.8 < \alpha < -2.0$, we created artificial
datasets which might be collected by our survey.  This was done for
each simulated luminosity function by
(1) populating a pencil beam survey volume equivalent to the DMS
survey with galaxies; (2) ``detecting'' the galaxies in the volume,
given the apparent magnitude limit
and known redshift distribution
of our survey; (3) adding the redshift 
uncertainty of our photometric technique ($\sigma_z \simeq$ 0.05) to
each galaxy; and (4) computing an ``observed'' LF using our modified
1/V$_{max}$ method.  For each combination of $M*$, $\phi*$ and $\alpha$,
we simulated 100 datasets that each detected approximately 545
galaxies with Sbc colors or bluer, and used standard
bootstrap methods to estimate the expected spread of resultant
luminosity functions.  

We illustrate the results of these simulations with two examples.
In Figure 3, the dotted line represents the LF of the original 
parent population of galaxies, while the data points represent
the resultant ``observed'' LF.  The parent population was selected 
from a luminosity function with Schechter parameters $M* = -19.3$,
$\phi* = 0.0245$, and $\alpha = -1.2$; this is very close to the
LF derived from the Autofib Redshift Survey for $0.02 < z < 0.15$
(Ellis et al. 1996).  We have computed the resultant ``observed''
luminosity functions in three redshift bins: $0.02 \leq z < 0.15$,
$0.15 \leq z < 0.3$ and $0.3 \leq z \leq 0.5$.  The systematic
effects are clearly evident, and differential with redshift as well;
the low-redshift LF appears to have a steeper faint-end slope,
while the medium- and high-redshift LFs seem to have
much higher M*.  For all three LFs, $\phi*$ is observed
to be lower than the parent population's original value.

The results of a similar set of simulations, but with $\alpha = -1.8$,
are presented in Figure 4.  This time, although the M* and $\phi*$
effects are qualitatively similar to those in Figure 3, the 
``observed'' results reflect $\alpha$ much more accurately in the
low-redshift bin.  (In the medium- and high-redshift bins, not
enough faint galaxies are observed to make an accurate
measurement of $\alpha$; this is consistent with the actual
survey data as well.)  The reduced distortion in $\alpha$ with
increasingly steep LFs is consistent with the analysis of
fuzzy redshift LFs in SubbaRao et al. (1996).

Our simulations show that the numerical effects of the fuzzy
redshift, although sometimes severe, are predictable and repeatable.
This means that the luminosity functions in each redshift bin can be
properly and unambiguously corrected to reflect 
more accurately the original parent luminosity function.  The
amount of correction needed varies depending on the parameters of
the parent LF, and the M* and $\phi*$ determined after corrections
are applied are still uncertain at about the 25\% level.  However,
the faint-end slope of the luminosity function is very accurately
reproduced ($\Delta\alpha = \pm0.05$) as long as the data extend
to at least $ M\simeq -16$.

Using the corrections computed with our simulations, we correct
the LF measurements in each magnitude bin, within each redshift bin.
We then compute the U-band luminosity functions for our blue
galaxy sample in the three redshift bins mentioned above:
low- (0.02 $\leq z <$ 0.15), medium- (0.15 $\leq z <$ 0.30),
and high-redshift (0.30 $\leq z \leq$ 0.50).  
These bins correspond approximately
with the redshift bins adopted by other authors for ease of comparison.
Coincidentally, they also suit the Sbc-and-bluer 
galaxy sample very naturally;
there are similar numbers of objects within each bin (161, 181 and 195
galaxies respectively), and there are only 8 objects in the sample
with $z > 0.5$.

\subsection{The U-band Luminosity Function vs. Redshift}

We present the corrected U-band luminosity functions for the
blue galaxy sample in Figure 5.
The $1\sigma$ error contours for the low-redshift LF are
presented in Figure 6.  

Using the standard Schechter (1976) parameterization,
the faint end slope of the low-redshift LF is best fit with a power
law $\alpha = -1.85 \pm 0.15$.  This slope is very steep 
compared to the derived Schechter-parameterized $\alpha$'s 
of $\sim 0.9 \pm 0.1$
found in the local blue galaxy luminosity function 
measurements of Lin et al. (1996),
Marzke et al. (1994), Loveday et al. (1992) and others;
it is also steeper than the values of 
$\alpha =$1.25 to 1.44 measured by
Ellis et al. (1996) and Heyl et al. (1997) for various
emission-line and blue galaxy subsets in the Autofib redshift survey.

Interestingly, we note that Marzke et al. (1994) measured the
luminosity function for morphologically selected Sm-Im
galaxies in the CfA Redshift Survey, and derived a value 
of $\alpha$ = $-$1.87.  
The fact that the U-band LF faint-end slope is so similar to that
of local Magellanic spirals and irregulars suggests that
we are seeing objects with spectrophotometric properties typical
of Magellanic galaxies -- that is, blue and actively forming stars.

Aside from color selection, we may also be selecting more
low-surface brightness galaxies than a typical one or two-passband
sample; by using any three of six passbands to identify a galaxy,
selecting an aperture size with a redder passband, and then measuring
all the U-band flux in that relatively large aperture, we may be allowing
more objects with low U-band surface brightness into the sample.
(Our central surface brightness limit is $\mu_{\rm o}$(U) $= 22.6$ 
magnitudes per square arcsecond.)
Since low-luminosity, low-surface brightness galaxies
tend to be bluer and may be
quite numerous (De Jong 1995; McGaugh, Bothun \& Schombert 1995),
and Magellanic galaxies tend also to have lower surface brightness,
the similarly steep slope of our LF and the Marzke et al. (1994) Sm-Im LF
may be due in part to a less severe surface brightness selection effect
in these samples compared to other samples used to compute LFs.

We note also that the surveys of
faint galaxies in clusters at $z \sim 0.2$ conducted by
Smith, Driver \& Phillips (1997) and Wilson et al. (1997)
have found steep $\alpha$'s of $-1.7$ to $-2$ in those
galaxy populations as well.  Wilson et al. further suggest
that many of the fainter galaxies in their sample may have faded
significantly in surface brightness since $z=0.2$; and Smith
et al. speculate that such steep faint-end slopes may hold
for field galaxies as well (see also Driver \& Phillips 1996).  
It is an intriguing possibility
that the U-selected, blue field galaxy LF is drawn from the same
population as faint galaxies in rich clusters; their very similar
faint-end slopes lend some credence to this hypothesis.

\subsection{The U-band Luminosity Function From $z$ = 0.15 to 0.50}

In our medium-redshift and high-redshift bins, the U-band apparent 
magnitude limit allows us to sample only relatively bright galaxies
(M(U) $\sim -$18 or brighter). Schechter function fits to these
bright-end segments of the LFs, therefore, do not 
constrain $\alpha$ very rigorously.  We can, however, measure the relative
changes in M* and $\phi$* by fixing the LF segments to $\alpha$'s
consistent with the respective redshift bins, and finding 
the best fit to the other two parameters.  

The fits to the data were all obtained using an error-weighted
least squares algorithm.  
Good fits were possible for both segments
in the range $-$1.4$ \leq \alpha \leq -$1.8; so we
adopted the value in the center of that range, $\alpha$ = $-$1.6.  
Next, the best fit for the medium-redshift bin, with M* and $\phi$* as
free parameters, was computed with fixed $\alpha$.
(We emphasize once more that
these values are not meant to be ``measurements'' of these
parameters.  Our intent is to use them as benchmarks, to quantify
our comparison of the medium and high-redshift LFs.)  Finally,
we started from these values and moved M* and $\phi$* 
until a good fit to the data was achieved with the high-redshift LF 
segment.  The faintest luminosity bin in the 
high-redshift LF segment is
affected by incompleteness, and we do not include it in the fit.
Since the high-redshift segments barely reach M*, $\phi$*
can easily be varied as much as a factor of 50\% or more (with a
smaller corresponding move of M*) and excellent fits can still be
achieved; we thus chose the smallest possible change in $\phi$* that
produced a good fit to the data.  In this way, we can give a lower
bound to the number density evolution we might expect as the average
redshift of our sample increases from $<z> \sim $0.2 to $<z> \sim $0.4.

The best-fit Schechter function fit for the medium-redshift bin
was M*$= -19.61$ and $\phi* = 0.0147$.  With the prescription 
described above, the high-redshift bin was best fit with M*$= -19.85$
and $\phi* = 0.0187$.  In other words,
$\phi$* increases 30\% and M* brightens by 0.25 magnitude.
The formal errors to these fits are relatively large -- $\sim 0.2$ 
magnitude in M*, $\sim$25\% in $\phi$* -- but the trend 
is clear, even upon visual
inspection of Figure 5.  Our fitting procedure, which attempts to produce
the smallest possible shift in $\phi$*, also supports the idea that
we are seeing real evolution in the galaxy population.

Comparison with the CFRS and Autofib data shows that
our results are consistent with the evolution observed in those surveys.
In CFRS VI, M* in their blue galaxy luminosity functions brightens by 
about 1 magnitude, assuming no change in $\phi$*, between their redshift bins
0.2 $< z <$ 0.5 and 0.5 $< z <$ 0.75.  Our brightening of 0.25 magnitude
occurs largely within their low-redshift bin; it can be increased to $\sim$0.5
magnitude if we fit the parts of our LF that correspond
to the same magnitude ranges observed in the CFRS and hold $\phi$* fixed.
Both these values are consistent with an incremental brightening of M*
from $<z>\sim $0.2 to $<z>\sim $0.4, which leads to a total brightening
of $\sim$1 magnitude by $<z> \sim $0.6-0.7.  Heyl et al. (1997) parameterized
the Autofib data with redshift-dependent evolving luminosity functions as
a function of galaxy spectral type (determined by cross-correlation of
spectral features with galaxy templates); based on those models,
we can estimate the expected evolution in the
combined population of Sbc, Scd and Sdm/starburst spectral types
from $<z>\sim$0.2 to $<z>\sim$0.4.
For that interval, M* would brighten by $\sim$0.1 magnitude and $\phi$* would 
increase $\sim$50\%.
This solution is also allowed by our fits, which as we mentioned above
can easily accommodate an increase in $\phi$* by reducing $\Delta $M*.

\section{An Excess Starbursting Population At z$>$0.3}

All of our measurements of the evolution of the blue galaxy LF with redshift
are consistent with those observed by other workers using deeper
surveys.  We can now use the added dimension of multicolor observations in our
survey to examine the spectrophotometric properties of the galaxies themselves,
and see which kinds of galaxies are contributing most to the evolution.

We can glance at the distribution of color vs. magnitude for the entire 
blue galaxy sample in Figure 7.  
We plot rest frame (U-B) vs. absolute U magnitude,
with different symbols representing the different redshift bins.
The majority of the sample objects lie around (U-B) $\sim -0.1$,
as expected for a population dominated by late-type spirals and
starburst galaxies.  The reddest galaxies in the sample have (U-B) $\sim 0.2$;
this is determined by our blue galaxy selection criterion.
As (U-B) decreases, a progressively stronger young stellar population 
dominates the luminosity, implying increasingly active star formation.
A dotted line at (U-B) $= -0.35$ is provided in the figure, to mark
the approximate color of a galaxy with an ongoing global starburst.

Below this line at (U-B) $= -0.35$, 
objects from the high-redshift bin
outnumbers objects from the lower-redshift 
bins by a factor of two.
At first glance this may not seem surprising, since the high-redshift bin
samples a comoving volume about twice that of the two lower redshift bins
combined.  The respective fractions of these objects in their bins, however,
is significantly different: less than 4\% for the lower redshift bins,
compared to 10\% at high redshift.  The difference cannot be
attributed to random errors alone.

A quantitative display of this effect is shown in Figure 8.  The galaxies
have been placed into nine bins -- 3 U-band luminosity bins, in each of
the three redshift bins -- and plotted in histogram form.  (Bins containing
no galaxies are plotted for reference.)
In the vertical direction, going down,
the galaxies in the same redshift bin increase in luminosity.  Down each
column, in each
redshift, we see the well known color-magnitude relation, where brighter
galaxies tend to have redder colors.  But across the rows,
the magnitude bins remain constant, and only redshift changes; although
the number of galaxies in each bin varies widely, the color
evolution with redshift is evident.  For example, the median (U-B) color
of the high-redshift bin from $-19.1<$M(U)$< -21.1$ is 
about 0.1 magnitude bluer than those at medium-redshift, 
and about 0.2 magnitude bluer than those at low-redshift.
The trend exists regardless of what boundaries for
the luminosity bins are chosen.

We test the possibility that this is a magnitude-limited selection effect
by plotting the same figure with a brighter limiting magnitude of
U$\leq$20.8 instead of U$\leq$21.1, and present it in Figure 9.  The
number of galaxies is reduced, so the statistics are less reliable.
Nonetheless, the same effect is observed, and the median colors for
each bin are essentially unchanged.  

As a further check on the excess of very blue galaxies
in the high-redshift bin, we examine the U, B, and R
images of these galaxies in the survey data.  Each galaxy
appears to have an atypically bright U-band image, with no
signs of undiscovered cosmic rays, uncorrected bad pixels, or other
problems in the data that would have artificially caused the 
colors to be so blue.  
If they existed at low redshift, these objects would easily
have been detected --- and even preferentially selected --- in our 
U-selected survey.  But there are almost no such galaxies
in bins (d) and (g) of Figures 8 and 9.  
The implication seems to be that a 
significant population of very blue, probably starbursting galaxies
appears at $z\gtrsim$0.3 which are not observed at $z<$0.3.
%should be noted: first, blue galaxies at
%low redshift should be preferentially selected in the 

\subsection{Spectroscopic Confirmation}

There remains the possibility that these very blue objects are not
actually galaxies at the redshifts we believe, but rather 
artifacts caused by incorrectly
determined photometric redshifts.  We thus use spectroscopy to confirm
that this excess starbursting population exists.

Spectra were obtained using the the Boller \& Chivens Spectrograph
on the Steward Observatory 2.3-meter telescope at Kitt Peak on 
December 1996 and January 1997.  We used a 4$\arcsec$.5 $\times$ 180$\arcsec$
longslit and a 400 l/mm grating, to obtain spectra from 3650 \AA\ to 
7000 \AA . By using the
large slit width, we were able to obtain essentially integrated spectra
of the objects and avoid problems with
differential atmospheric refraction.  Spectral resolution was
usually not limited by the slit width, and was typically $\sim$15 \AA .
We also observed  spectrophotometric standard stars
to obtain relative spectrophotometry for the galaxies.
Standard data reduction techniques using the IRAF software system
were used to process the longslit data,
primarily with the CCDRED and LONGSLIT packages.
Apertures along the slit were traced and extracted
with the APEXTRACT package.  

The apparent magnitudes and integration times of the
objects we observed are given in Table 1.  Usually, it is rather
difficult to obtain spectra for 20th magnitude or fainter objects with
a 2-meter class telescope; but the high sensitivity of the Steward 800x1200
CCD, combined with the strong emission lines of the targets, allowed us
to obtain unambiguous redshifts for all 9 blue objects we observed.
The color redshift determinations and spectroscopic measurements are also
given in Table 1.  The mean absolute deviation of the photometric vs.
spectroscopic redshifts is $\Delta z$ = 0.043, exactly consistent
with the accuracy level we predict for our redshift estimation algorithm
(LG98).

Four of the spectra are shown in Figure 10.  We overplot the redshifted
spectral energy distribution of our starburst template 
over each spectrum {\it (dotted line)} 
for comparison; the continua for both the templates and the
galaxies have been normalized to unity at the [OII]$\lambda$3727 \AA\ 
emission line.   The spectra show that these galaxies are 
indeed blue galaxies similar to the template starburst galaxy.
This further confirms that our observed excess starburst population is real.

\section{Discussion}

The idea of a bursting population at relatively high redshift that has
since faded has been put forth, among others, by Broadhurst, Ellis \& 
Shanks (1988), Lacey \& Silk (1991), Babul \& Rees (1992), 
and Cowie et al. (1996).  The
application of those models, however, is primarily to the so-called 
``faint blue galaxies,'' (see, e.g., Ellis 1997)
where galaxy counts beyond B$\sim$22 are much
greater than expected for a non-evolving population.  
Our blue galaxy survey from the DMS only reaches B$\sim$21; thus
it is probably inappropriate to call the excess starburst galaxies in
our high-redshift bin ``faint blue galaxies'' in the above sense.
However, our data certainly suggest that our starbursts may be 
related to that faint blue galaxy population, and in fact may be 
the lowest redshift examples of those distant and numerous blue objects.

What these $z \gtrsim $0.3 high redshift starburst galaxies may help explain 
is the general evolution of the blue galaxy luminosity function.
We have confirmed the results of Koo (1986), 
CFRS VI, Ellis et al. (1996), Heyl et al. (1997), Lin et al. (1997)
and others that evolution does occur
in the blue field galaxy population, and that the evolution is
observationally discernible by $z\simeq$0.3.  The 
number density and/or luminosity increases we are seeing 
may in fact be caused in part by the appearance of these starbursting objects.
No galaxy can long sustain the gas consumption rate required to
produce the strong global star formation implied by their (U$-$R)
and (U$-$B) colors;
so these objects should eventually fade, then redden,
and eventually blend into the more numerous population
of galaxies with lower, steady-state star formation.

Can such starburst galaxies fade into the background of more quiescent galaxies
in the time between our high-redshift and medium-redshift bins?  We examine
this question using the population synthesis models of Bruzual \& Charlot
(1993),
which give us quantitative estimates of the colors and magnitudes of
aging starbursts.  We first consider the most extreme case, of a galaxy
starbursting so strongly that its underlying stellar population contributes
negligibly to the total luminosity of the galaxy.  For this case, we use
an instantaneous burst model with a Salpeter initial mass function.  In
our chosen cosmology, 1.4 $\times$ 10$^9$ years elapse between $z$=0.4
and $z$=0.2, roughly the mean redshift of the galaxies in our high-redshift
and low-redshift bins respectively.  We assume that we are observing the
burst at an age where the rest frame (U-B) and (U-R) colors most closely
match those of the starburst galaxies in the high-redshift bin.  That
age would be 6 $\times$ 10$^7$ years, when (U-B) = $-$0.49  and 
(U-R) = 0.28.  After 1.4 $\times$ 10$^9$ yr the U magnitude of the burst
will have faded over 4 magnitudes according to the Bruzual \& Charlot
models, and the colors
will have reddened to (U-B) = 0.24 and (U-R) = 1.3.  If M(U)$ \sim -$21
at 6 $\times$ 10$^7$ yr, the post-burst evolution moves the galaxy
well into the middle of the locus where most of the low-redshift galaxies lie
in Figure 7.  Our survey would not be able to detect such an 
object at $z \sim$ 0.2.

A somewhat less extreme case would be a galaxy which has a global
starburst triggered in it.  In one possible scenario, explored by 
Charlot \& Silk (1994), Belloni 
et al. (1995), Barger et al. (1996) and others, the starburst 
continues at a constant star
formation rate for a short time (typically 10$^8$ to 10$^9$ yr), 
then stops after converting into stars a gas fraction
equal to some percentage (typically 10-20\%) of the final mass of the galaxy.
After this burst is over, all star formation is truncated.
We choose for our comparison a late-type spiral galaxy which undergoes a
Salpeter-IMF starburst lasting 10$^8$ yr 
that consumes 10\% of the final galaxy mass.  If we select a time after
the starburst begins when the galaxy has UBR colors similar to our
template starburst, these authors show that (U-B) will again be 
$\gtrsim$0.2 after 1.4 Gyr; 
the U-band fading would be a more modest $\sim$2 magnitudes.
This burst scenario would move our M(U)$ \sim -$21 bursting galaxy into
the middle of the medium-redshift galaxy locus, again 
becoming largely anonymous within a large reservoir 
of ordinarily-colored blue galaxies.

If fading and reddening of the burst
is indeed the mechanism for removing
starburst galaxies from view nearward of z$\sim$0.3, one strong test
of this population's effect on the blue galaxy luminosity function would
be to see if removing the excess starburst galaxies leads to a luminosity
function consistent with a passively evolving galaxy population since
$<z> \sim$ 0.4.  Our photometry is unfortunately 
not deep enough to let us conduct this test rigorously.  
A firm conclusion can be drawn when additional data for
galaxies at fainter absolute magnitudes become available.

\section{Conclusions}

We have shown that an optical multicolor survey of field galaxies,
such as the Deep Multicolor Survey, can be a very powerful tool for studying
galaxy evolution.  Such a survey offers concrete advantages over surveys
with only one or two passbands: the two most relevant in this work
are our decreased dependence on K-corrections
for accurate absolute magnitude determinations, and the additional
leverage we obtain from examining the spectral energy distributions
of each galaxy.  In addition, while there is ultimately no substitute 
for secure spectroscopic redshifts, we can
extract almost as much information about luminosity functions and 
evolution as a function of redshift as true redshift surveys by using a
photometric redshift-classification technique such as we have, with
orders of magnitude less telescope time.  
We have developed a modified version of the 1/V$_{max}$ method for
computing luminosity functions using galaxies with photometrically
determined redshifts.  The method's systematic errors make absolute
determinations of M* and $\phi*$ difficult; but relative changes
in the luminosity functions are reliably measured, and can be 
effectively used to study differential evolution.

We have assembled a complete, magnitude limited sample of 545 galaxies
with rest-frame multicolors as blue as, or bluer than, a typical Sbc galaxy.
The low-redshift (0.02 $\leq z <$ 0.15) luminosity 
function for this sample has a very steep faint-end slope, which turns out
to be consistent with the measurement of $\alpha$ for Magellanic
spirals and irregulars from the CfA Redshift Survey (Marzke et al. 1994).
The implication is that our blue galaxies and those Sm-Im galaxies are
drawn from essentially the same steep-sloped  population.  Whether
that population is defined by its spectrophotometric, morphological or
surface-brightness properties is uncertain, and merits further 
investigation.

U-band number counts vs. redshift, and comparison of the luminosity
function segments in medium-redshift (0.15 $\leq z <$ 0.30) and
high-redshift (0.30 $\leq z \leq$ 0.50) bins, demonstrate significant
evolution in the galaxy population which is clearly 
visible by $z \gtrsim$ 0.3.
The nature, amplitude and epoch of the evolution we observe are 
consistent with those found 
by the Canada-France Redshift Survey and the Autofib redshift survey.
Using the broad wavelength coverage for each galaxy in our survey, we
use color-magnitude diagrams and histograms in (U$-$B) vs.
absolute magnitude to identify an excess population of apparently
starbursting galaxies in the high-redshift bin which does not appear
at lower redshift.  It is plausible that
these objects have been temporarily brightened by their global starbursts,
and will redden and fade into obscurity by $z \sim$ 0.2.
These galaxies may be contributing significantly to the observed evolution
of the blue galaxy luminosity function at $z \gtrsim$ 0.3.  If this is
true, what makes the particular epoch $z \simeq$ 0.3 the threshold past
which these starbursts are no longer produced?  The answer to that question
will contribute greatly to our understanding of field galaxy evolution.

\acknowledgments
 
We thank Rob Kennicutt, Jim Liebert, Hans-Walter Rix and Jon Gardner for
helpful discussions.  We thank the referee for a detailed and very
constructive critique which has significantly improved this work.
%C. L. gratefully acknowledges support from NASA grant NGT-50758. 

\appendix
\section{Appendix I \\ 
The Deep Multicolor Survey: Galaxy Detection, Photometry, and
Redshift-Type Identification}

A detailed description of the Deep Multicolor Survey (DMS) is given
in HOGPW.  Here, we summarize its 
characteristics, and describe how the galaxy sample is derived.

The DMS was obtained with the Mayall 4-meter telescope at KPNO, in
direct imaging mode at prime focus, with an engineering quality 
2048$\times$2048 Tektronix CCD.  The survey covers 0.83 deg$^2$
along six lines of sight at high galactic latitude.  Each field
was observed with six filters: standard Johnson UBV; a custom
R filter calibrated to the Kron-Cousins system; and two custom
I filters with $\lambda_{eff}$ = 7430 \AA\ and 8520 \AA\
respectively, referred to hereafter as I75 and I86. Reduction and
calibration of the images, and the establishment of the photometric
system for the nonstandard filters, was performed as described in HOGPW.

The assembly of the object catalog was performed in several steps.
First, the Faint Object Classification and Analysis System (FOCAS; Valdes
1982a) was used to identify objects using its default (``built-in'') 
detection filter.  The detected objects were classified using the
``resolution'' task in FOCAS (Valdes 1982b) as star, fuzzy star,
galaxy, diffuse object, or noise, using templates generated from
a point source function empirically determined from the CCD image.
An inclusive, automatically generated galaxy catalog was then assembled
using all the objects classified as galaxies in at least three
filters.  In all but one of the surveyed lines of sight, two exposures
were taken in each passband; in those cases, ``resolution''
had to classify an object as a galaxy in both exposures for it to be
declared a galaxy in that passband.  This inclusive catalog contained
9,431 objects.

The second step was to use the IRAF package APPHOT to obtain aperture
photometry of each object in the catalog described above.  Each
galaxy's flux was measured with concentric circular apertures 
ranging from 10 to 30 pixels (5$\arcsec$.3 to 15$\arcsec$.9) in
diameter. Instrumental magnitudes and Poisson signal-to-noise
were measured for each aperture; the sky value was computed
separately for each object by taking the mode of the pixel
values in an annulus around the aperture center, typically with
inner diameter 32 pixels 
(17$\arcsec$.0) and outer diameter 50 pixels (26$\arcsec$.5). 

An aperture optimization technique,
similar to the growth curve optimization method used 
by Yee, Green \& Stockman (1986),
%%REF Yee, H. K. C., Green, R.F., \& Stockman, H. S. 1986, \apjs, 62, 681.
was then applied to each object.  The function of the object's Poisson
signal-to-noise vs. aperture radius was examined and the optimal
size was determined to be either: (1) at the 30 pixel diameter 
(15$\arcsec$.9) limit; or (2) where the next largest aperture 
showed an inflection point, indicating an intruding object
or cosmic ray, or a decrease much larger than expected from the
addition of random sky noise.
The aperture size selection was performed using the R-band data,
the passband with the greatest depth, for each object; extinction
and color-corrected magnitudes
were then extracted with the same aperture size in all six passbands.
%We estimate that this relatively crude method for obtaining  ``first cut''
%photometry for all the survey objects contains systematic errors
%of at most 10\%.

%and to check that our overall galaxy number densities were
%consistent with those in the literature.
The U-band magnitude selection limit for our blue galaxy sample
was determined by our desire to
have a complete sample with good signal-to-noise in all passbands
for each galaxy.  In the DMS, this meant a typical magnitude of
U $\lesssim$ 21.1 for a galaxy with surface brightness levels
typical of a Magellanic spiral.  
Visual inspection of the raw number counts,
presented in Figure 1, shows that 
this is over a magnitude brighter than the U-band completeness limit
of U$_{lim} \sim$ 22.2; thus we are confident that such a selection
limit would also be a complete, magnitude-limited sample.

To make sure all the galaxies within our magnitude limit would be
included, we took the subset of all the objects in the inclusive
survey brighter than U=21.2 -- the magnitude limit desired, plus
the typical 1$\sigma$ photometric error at that limit -- and 
inspected them visually.  Using the IMEXAM task in IRAF, we checked
that the automatically optimized aperture for each galaxy was indeed
appropriate -- that is, inclusive of all the galaxy light within
the sky-limited isophotal magnitude, and not inclusive of nearby
objects.  Objects with stars or very large ($>$6 contiguous pixels) cosmic
rays within the radius of our minimum aperture size (2$\arcsec$.6)
were removed from the sample; these objects comprised about 5\%
(44 out of 953) of the original subsample.
In cases where the aperture optimization algorithm was ``deceived''
(e.g., by a highly edge-on galaxy, or a strongly interacting system),
the aperture size was adjusted appropriately.  

The instrumental fluxes for the visually adjusted apertures were
then extracted from the photometry database.
Smaller, visually obvious cosmic rays in or near the apertures
that were not removed in the initial calibration
and reduction of the image frames (see HOGPW) were removed by
hand using the FIXPIX task in IRAF, which replaces the affected
pixels with a value interpolated from the surrounding pixels.
Extinction and color-corrected magnitudes were again computed.
In the lines of
sight where two frames were available for each passband, a final
additional check was made for cosmic rays and bad pixels: 
if the signal-to-noise
of one image in a given filter was more than 150\% of the other image,
it was assumed that a cosmic ray, chip defect or other systematic error
was contaminating that image, and the uninflated measurement was used.
Otherwise, the final magnitude for each bandpass was computed
as a signal-to-noise weighted average of the two measurements.
All the objects with 17.0$\leq$U$\leq$21.1 were then extracted.  This
subsample contains 667 objects.

Finally, every object in the sample was processed using the 
``GetZ'' program as described in LG98.  
``GetZ'' implements the photometric 
redshift-spectral classification algorithm (see Section 3), and outputs 
a redshift and a spectral type (E/S0, Sab, Sbc, Scd, or Irr).
``GetZ'' did not find an acceptable solution to 8 objects;
inspection showed that all of these objects appeared to be
point sources or nearly so.  These objects were taken to be
misclassified stars, and removed from the sample.  There is a
small possibility that these objects were AGN, whose colors are
sufficiently different from the template galaxies that ``GetZ''
could not find redshift-type matches. Their removal in any case
should not affect our conclusions, since their numerical
contribution to the total sample is so small (barely 1\%).
The final sample contains 659 galaxies, each of which has 
$\sim$10\% or better photometry in all six passbands.

\clearpage

\begin{deluxetable}{lcccc}
\scriptsize
\tablewidth{0pc}
\tablecaption{Spectroscopic Results}
\tablehead{
\colhead{Galaxy ID}         & \colhead{mag(B)}     &
\colhead{Exposure}          &
\colhead{$z_{phot}$}        & \colhead{$z_{spec}$}
}

\startdata

01a-521-225  & 20.75 & 2300s  & 0.250  & 0.239  \nl %(2) 
01a-1085-740 & 20.54 & 2500s  & 0.250  & 0.245  \nl %(2) 
01a-632-1866 & 21.27 & 3600s  & 0.275  & 0.236  \nl %(2) 
10a-1762-311 & 20.07 & 3600s  & 0.300  & 0.109  \nl %(2) 
01a-246-1506 & 20.56 & 4800s  & 0.325  & 0.283  \nl %(2) 
10a-636-1395 & 21.41 & 2400s  & 0.325  & 0.296  \nl %(2) 
01a-1680-382 & 21.50 & 4000s  & 0.375  & 0.360  \nl %(3) 
01a-448-224  & 20.84 & 2300s  & 0.400  & 0.430  \nl %(2) 
01a-1029-680 & 20.94 & 3600s  & 0.475  & 0.492  \nl %(3) 

\enddata
\end{deluxetable}

\clearpage

%\centerline{{\bf FIGURE CAPTIONS}}

{\bf Fig. 1 } Total raw U-band galaxy number counts for the 
Deep Multicolor Survey. The dotted line at U=21.1 denotes the
magnitude limit for the U-selected sample.

\plotone{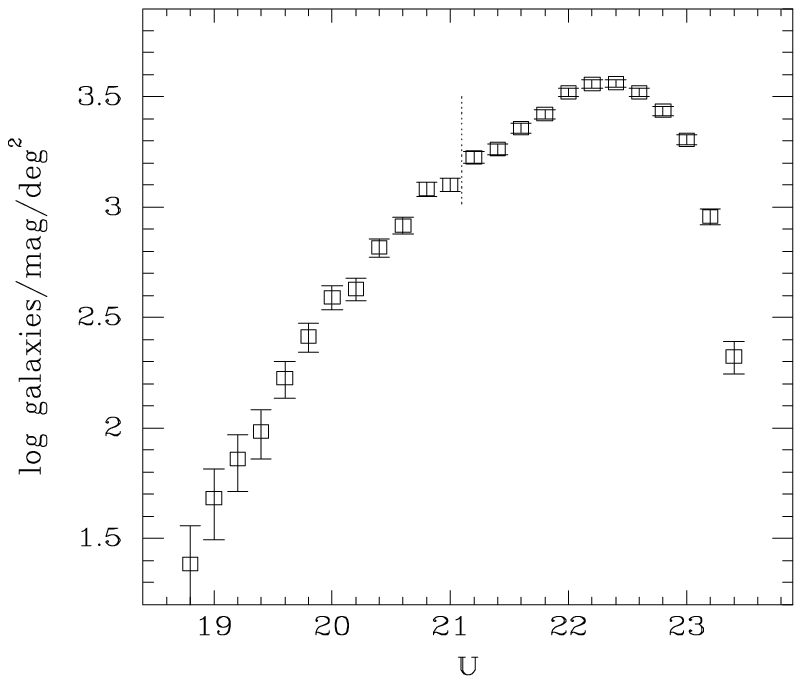}

\clearpage
%%%%%%%%%%%%%%%%%%%%%%%%%

{\bf Fig. 2 } U-band number counts in the U-selected sample
(dark circles).  Also plotted are the photographic U-band number
counts of Koo (1986) and Jones et al. (1991).

\plotone{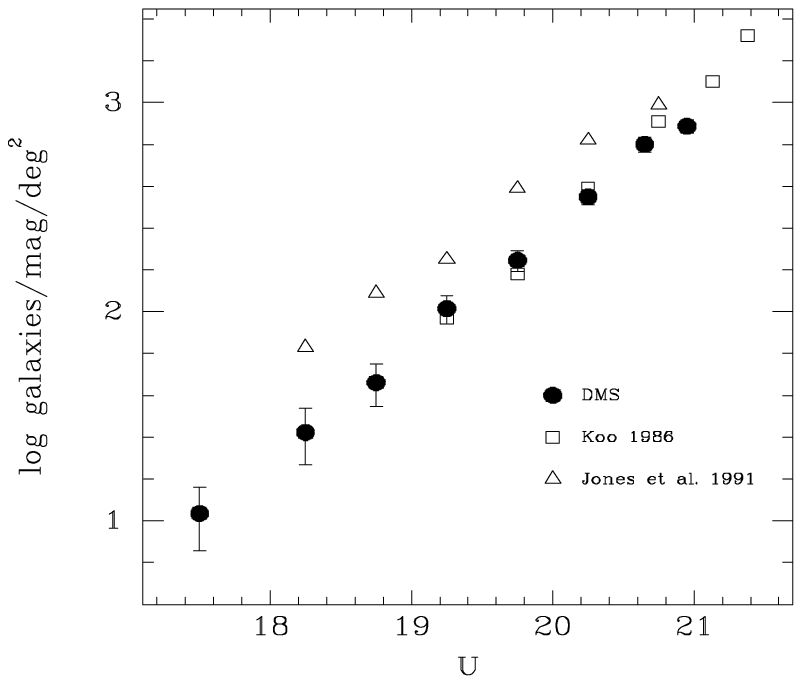}

\clearpage
%%%%%%%%%%%%%%%%%%%%%%%%%

{\bf Fig. 3 } Results of simulated observations using fuzzy redshifts.
{\it Solid line:} model luminosity function with M*$= -19.3$, 
$\phi* = 0.0245$ and $\alpha = -1.2$.  The points are the luminosity
functions recovered by the simulated observations for low, medium
and high redshift bins.

\plotone{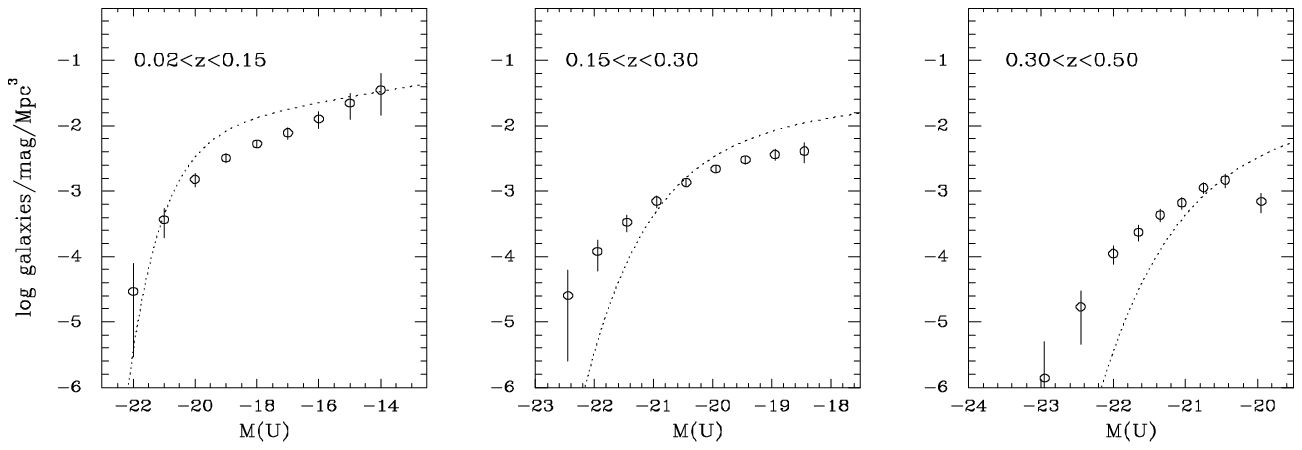}

\clearpage
%%%%%%%%%%%%%%%%%%%%%%%%%

{\bf Fig. 4 } Same as Fig. 3, but with $\alpha = - 1.8$.

\plotone{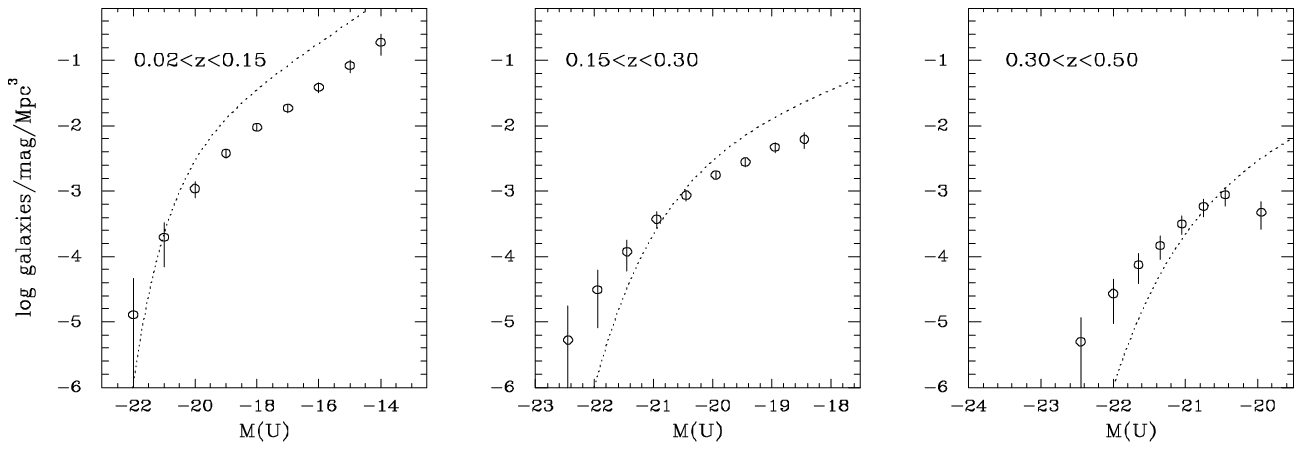}

\clearpage
%%%%%%%%%%%%%%%%%%%%%%%%%

{\bf Fig. 5 } Luminosity functions constructed from the probability-weighted
luminosity distributions of the blue (Sbc and bluer) galaxy sample.

\plotone{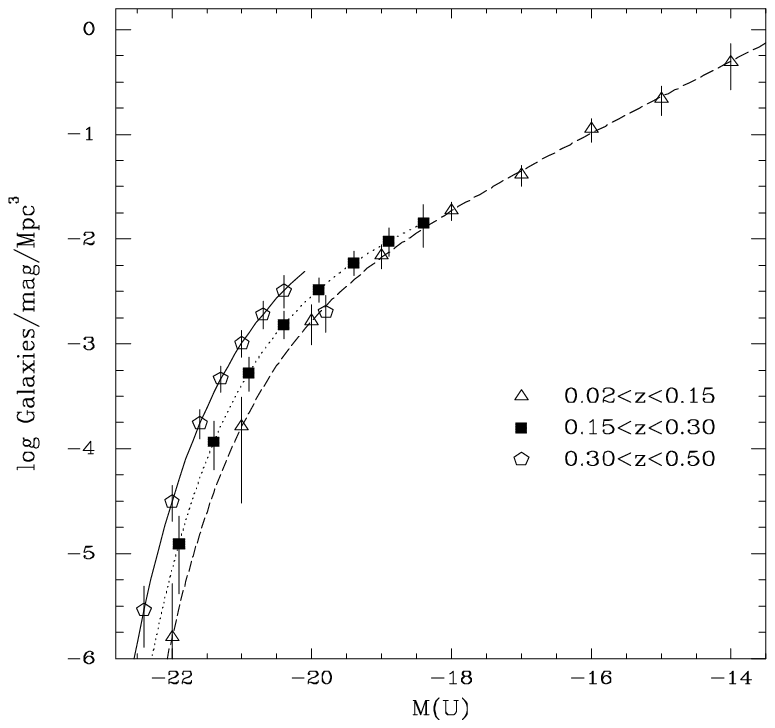}

\clearpage
%%%%%%%%%%%%%%%%%%%%%%%%%

{\bf Fig. 6 } 1$\sigma$ error contours for Schechter function fit parameters
for the low-redshift ($0.02 < z < 0.15$)
luminosity function of the blue galaxy sample.  
The point in the center of each contour
marks the Schechter function fit presented in Figure 3.

\plotone{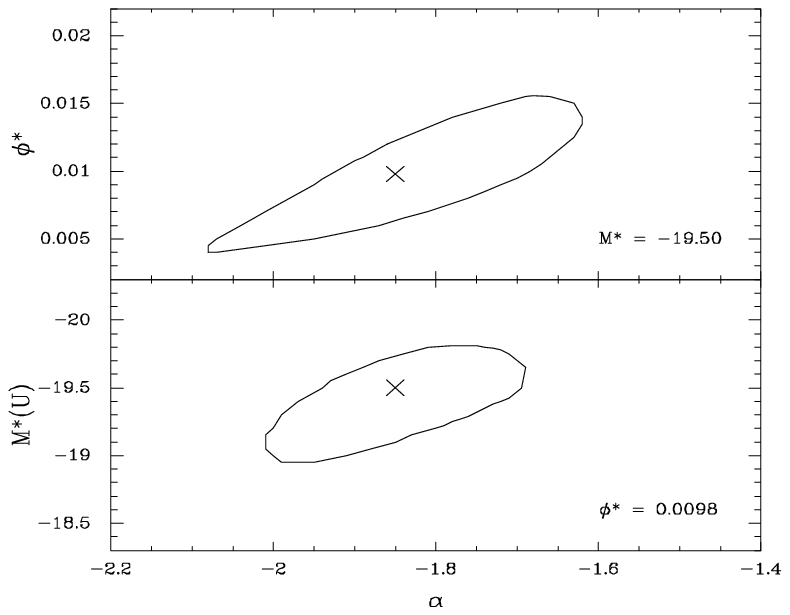}

\clearpage
%%%%%%%%%%%%%%%%%%%%%%%%%

{\bf Fig. 7 } Absolute R magnitude vs. rest frame (U$-$B) color for
the blue galaxy sample.  The symbol for each galaxy denotes its location
in the low (triangles), medium (squares) or high (circles) redshift bin.

\plotone{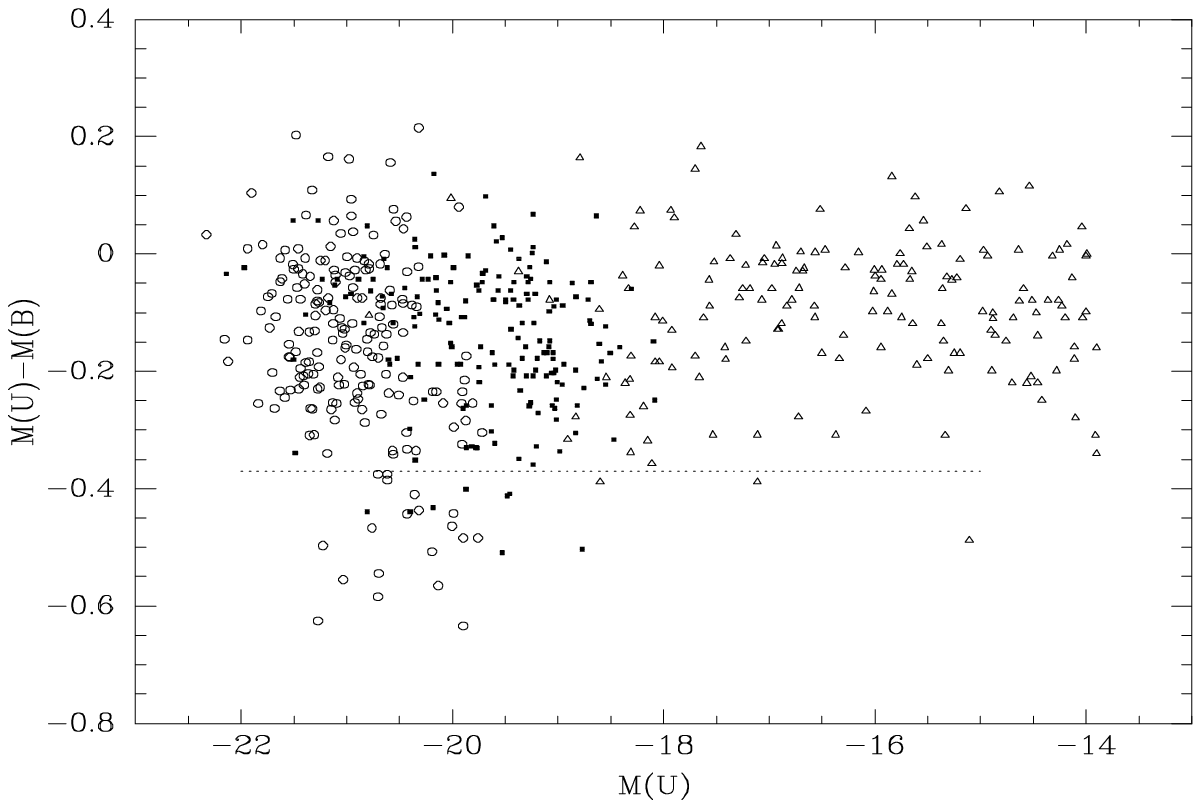}

\clearpage
%%%%%%%%%%%%%%%%%%%%%%%%%

{\bf Fig. 8 } Histograms of (U$-$B) vs. galaxy number for the blue galaxy
sample in three redshift bins and three absolute B magnitude bins.  
Each column contains galaxies in the same
redshift bin, while each row contains galaxies in the same magnitude bin.

\plotone{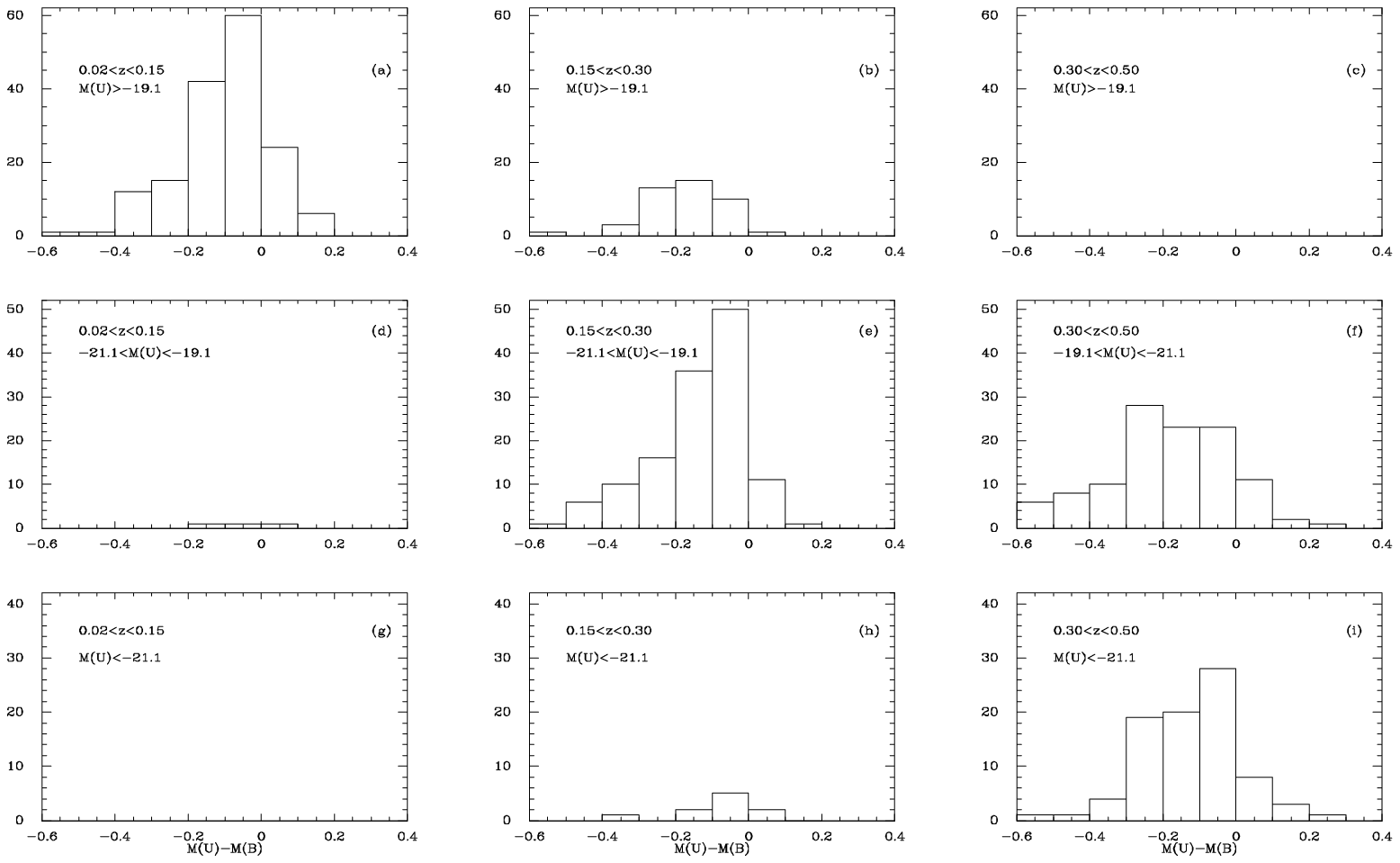}

\clearpage
%%%%%%%%%%%%%%%%%%%%%%%%%

{\bf Fig. 9 } Same as Figure 8, but with apparent magnitude limit reduced
to U $\leq$ 20.8 instead of 21.1.

\plotone{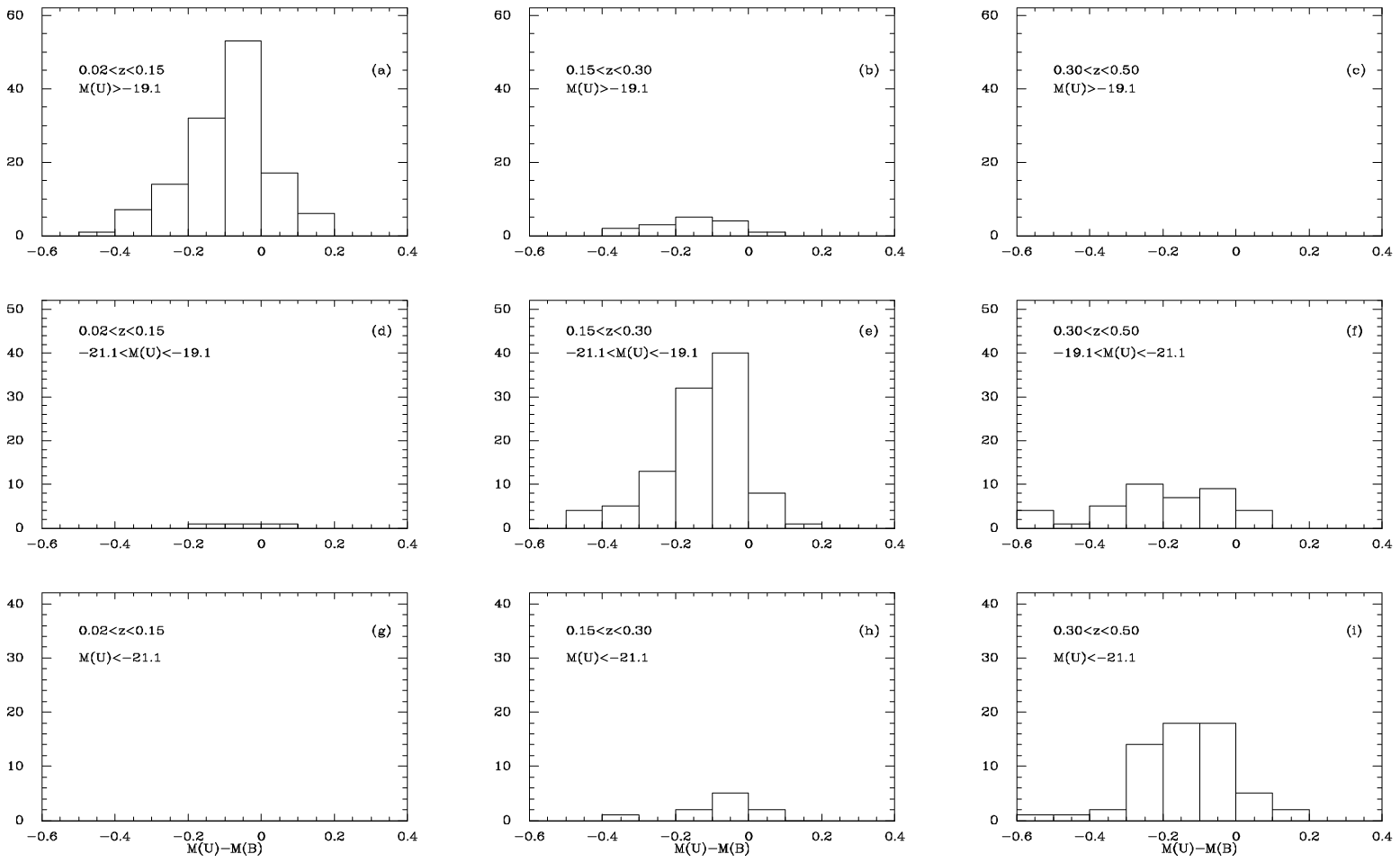}

\clearpage
%%%%%%%%%%%%%%%%%%%%%%%%%

{\bf Fig. 10 } Spectra of four of the excess blue galaxies.  Photometric
and spectroscopic redshifts for each galaxy are given.
The spectral energy distribution of the Liu \& Green (1998) starburst
galaxy template (dotted lines), redshifted to each galaxy, is
overplotted.  Spectra are in units of F$_{\lambda}$; wavelengths are
in \AA .

\plotone{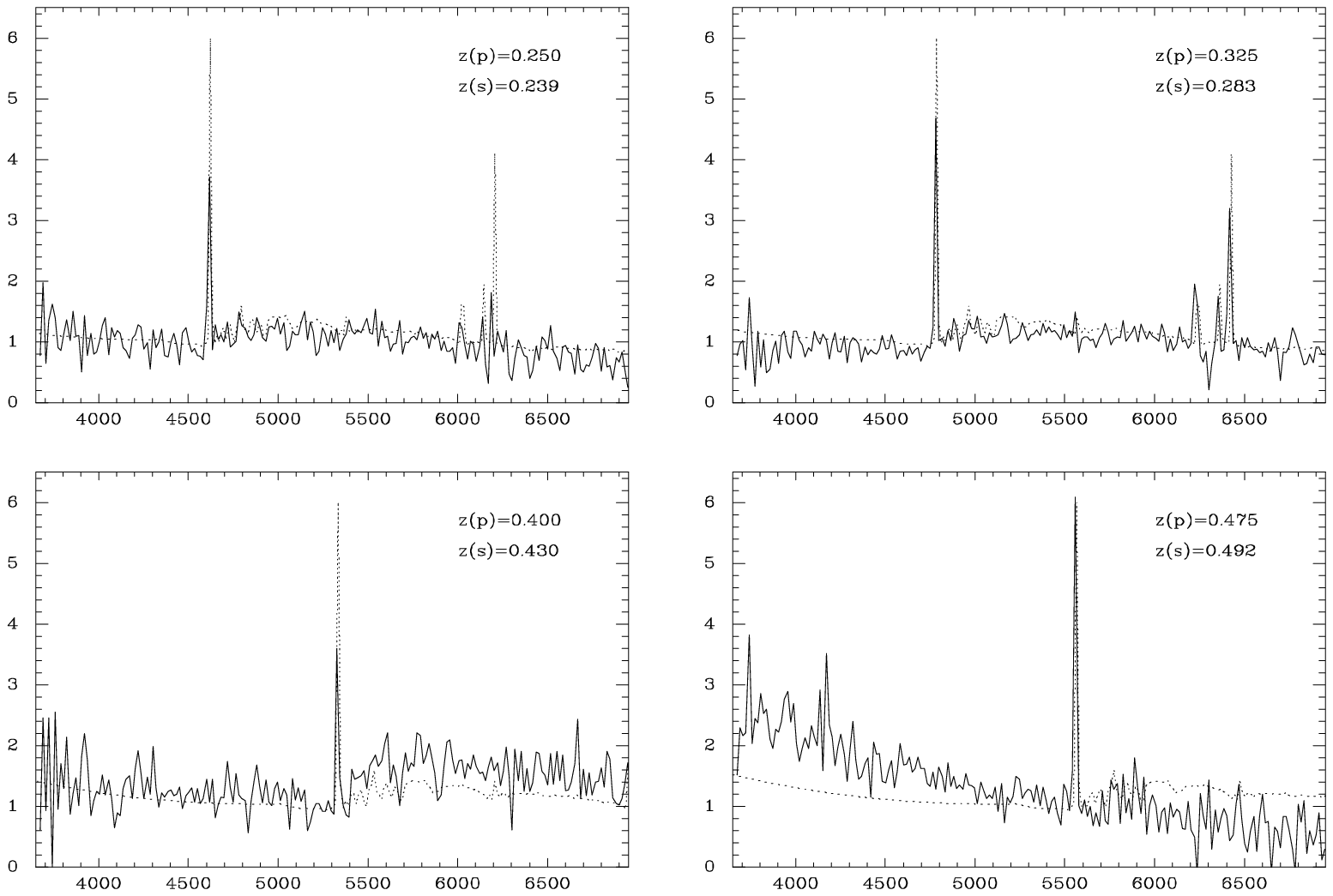}

\clearpage
%%%%%%%%%%%%%%%%%%%%%%%%%

\end{document}